# Probing individual layers in functional oxide multilayers by wavelength-dependent Raman scattering

J. Kreisel[1,2,3*], M.C. Weber[2], N. Dix[4], F. Sánchez[4], P.A. Thomas[2], J. Fontcuberta[4]

[1] Laboratoire des Matériaux et du Génie Physique, Grenoble INP, CNRS, Minatec, 3, parvis Louis Néel, 38016 Grenoble, France

[2] Department of Physics, University of Warwick, Coventry CV4 7AL, United Kingdom

[3] Science and Analysis of Materials, CRP Gabriel Lippmann, 41, rue du Brill, 4422 Belvaux, Luxembourg

[4] Institut de Ciència de Materials de Barcelona (ICMAB–CSIC), Campus UAB, Bellaterra 08193, Spain

* Corresponding author:   *jens.kreisel@grenoble-inp.fr*

## Abstract

Integration of functional oxides on silicon requires the use of complex heterostructures involving oxides of which the structure and properties strongly depend on the strain state and strain-mediated interface coupling. The experimental observation of strain-related effects of the individual components remains challenging. Here we report a Raman scattering investigation of complex multilayer $BaTiO_3/LaNiO_3/CeO_2/YSZ$ thin film structures on silicon. It is shown that the Raman signature of the multilayers differs significantly for three different laser wavelengths (633, 442 and 325 nm). Our results demonstrate that Raman scattering at various wavelengths allows both the identification of the individual layers of a functional oxide multilayers and monitoring their strain state. It is shown that all layers of the investigated multilayer are strained with respect to the bulk reference samples, and that strain induces a new crystal structure in the embedded $LaNiO_3$. Based on this, we demonstrate that Raman scattering at various wavelengths offers a well-adapted, non-destructive probe for the investigation of strain and structure changes, even in complex thin film heterostructures.

# 1. Introduction

Oxide thin films exhibit a remarkable range of magnetic, electrical, and optical properties and represent an area of increasing research interest in the field of functional materials. A considerable amount of current research goes to thin film oxide *multi*layers [1-4] - often also called superstructures, superlattices or heterostructures - which are defined as a sequence of thin film layers. The wide interest into oxide multilayers is triggered by various multilayer-specific characteristics: (*i*) Properties of multilayer may be superior to the parent materials, as illustrated by the reported enhancement of the dielectric constant in $BaTiO_3$-$SrTiO_3$-based multilayers.[5] (*ii*) Interface engineering can lead to new unexpected properties as exemplified by intriguing conducting electron systems at the interface between insulating $SrTiO_3$ and $LaAlO_3$ layers.[6-7] (*iii*) Multilayers are multifunctional materials *par excellence*, because they combine the properties of the individual layers as for instance in the combination of ferromagnetic and ferroelectric layers which lead to multiferroic materials.[8-10] (*iv*) The full integration of oxide thin films in current technologies, namely its growth on silicon together with suitable metallic layers for electrical connectivity, commonly requires complex multilayers.

In such complex stacking structures, each constituent layer (or buffer-layer) is usually only a few tens of nanometers thick which allows for full epitaxial growth of strained layers. Understanding the structure-property relationship of these heterostructures requires monitoring of the complex pattern of interface-mediated strain which can lead to both deformations within a space group or even structural phase transitions of the individual layers. Both effects are known to critically determine the properties of functional oxides. The observation of such strain-induced changes in complex thin film multilayers remains challenging. Specifically, their observation is difficult using diffraction techniques, particularly for those films with close or overlapping Bragg reflections of several different layers, consequently alternative non-destructive techniques are of great importance.

The aim of our work is to show that Raman scattering (RS) can provide some unique information when investigating the strain state in complex multilayers which consist of several different functional oxides. RS is a well-known non-destructive probe for investigating structural properties of oxides, namely with regard to subtle distortions and phase transitions.[11] After an early focus on bulk systems, RS is now also routinely applied for the characterization of thin oxide films.[12-17] By far most literature Raman scattering studies are conducted on simple thin films, while only some investigations concern repeated bilayers[14, 18-21] or complex pillar-matrix nanostructures.[22-23] The vast majority of Raman studies on oxide thin films use a single wavelength only, which is selected according to the best Raman signal or to reduce the substrate background signal and fluorescence.

In this report, we will demonstrate that the complementary use of different laser wavelengths allows the investigation of the individual layers and their strain state in complex functional multilayers. We argue that this approach is widely applicable and we will show that it allows discovering subtle structural modifications occurring in embedded layers. To illustrate the usefulness of this approach we have chosen to investigate a complex full-oxide heterostructure $BaTiO_3/LaNiO_3/CeO_2/YSZ/SiO_x//Si(001)$. This multilayer permits stabilization of the functional room-temperature ferroelectric $BaTiO_3$ (BTO) as fully epitaxial layers on Silicon with a large ferroelectric polarization pointing out-of-plane,[24-25] as is required for most common applications.[24-25] Integration of the metallic layer, $LaNiO_3$ (LNO) offers the benefit of balancing strain arising from structural mismatch, and the thermal strain arising from the expansion coefficients of Si and BTO, which are known to severely impact the BTO film stability and to impose undesired orientation of the polarization.[26-28]

## 2. Experimental

$BaTiO_3/LaNiO_3/CeO_2/YSZ$ epitaxial heterostructures were prepared by pulsed laser deposition in a single process on Si(001) substrates. A KrF excimer laser ($\lambda = 248$ nm) was focused sequentially on stoichiometric YSZ, $CeO_2$, $LaNiO_3$, and $BaTiO_3$ ceramic targets. The multilayer discussed here contain 40 nm, 20 nm, and 35 nm thick layers of YSZ, $CeO_2$, and $LaNiO_3$, respectively, and a 55 nm thick BTO top layer. Additional details about preparation conditions will be reported elsewhere.[29] The crystal orientation and out-of-plane lattice strain were investigated by X-ray diffraction using $CuK_\alpha$ radiation.

Raman spectra were recorded with a Renishaw inVia Reflex Raman Microscope. Experiments were conducted in micro-Raman mode at room temperature, using three different exciting wavelengths: a red 633 nm line from a HeNe laser; and two lines at 442 (blue) and 325 nm (UV) from a HeCd laser. The spectral cutoff for the visible 633 and 442 nm excitation lines is about 120 cm$^{-1}$, while the cut-off for the UV line is significantly higher, at about 550 cm$^{-1}$. Taking into account that Raman spectra of transition metal oxides often show a dependence on the exciting laser power due to overheating, we carefully checked several laser powers for the three wavelengths to ensure that our used laser power does not lead to a modified spectral signature. The reproducibility of spectra on different places of the sample has been also verified.

## 3. Structural data

The BaTiO$_3$/LaNiO$_3$/CeO$_2$/YSZ heterostructure is sketched in Figure 1.b. The bottom YSZ layer grows epitaxially with cube-on-cube epitaxial relationship on Si(001) without necessity of removing the native SiO$_x$.[30-31] The XRD θ/2θ scan around symmetrical reflections of the heterostructure is shown in Figure 1.a and zoomed in Figure 1.b. The high intensity peak corresponds to Si(004), and the others are (00*l*) reflections from YSZ, CeO$_2$, LaNiO$_3$, and BaTiO$_3$. Scans around asymmetrical reflections (not shown here) confirmed that all layers are epitaxial, with cube-on-cube growth of YSZ and CeO$_2$, and LaNiO$_3$ and BaTiO$_3$ rotated 45º in-plane respect to the Si lattice. YSZ and CeO$_2$ have out-of-plane lattice parameters of 5.16 Å and 5.41 Å, respectively, which are bulk-like values, whereas LaNiO$_3$ has a slightly compressed out-of-plane parameter (3.82 Å). Finally, the BaTiO$_3$ top layer has an expanded out-of-plane parameter with c = 4.069 Å (BTO bulk parameter, c = 4.038 Å).

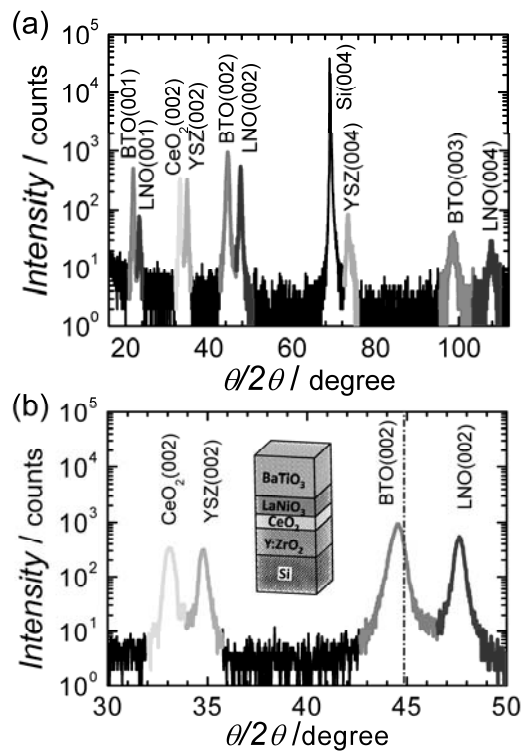

**Figure 1**
(a) θ-2θ XRD scan around symmetric reflections of the BaTiO$_3$/LaNiO$_3$/CeO$_2$/YSZ/Si(001) heterostructure sketched in (b) together with a zoomed region of the θ-2θ XRD pattern. BaTiO$_3$ and LaNiO$_3$ are labelled BTO and LNO, respectively.

## 4. Assignment of Raman modes in the multilayer

In the first step of our study, we aim at assigning the different modes in the Raman spectrum of the BaTiO$_3$/LaNiO$_3$/CeO$_2$/YSZ multilayer on a Si(001) substrate. Figure 2.a presents room-temperature Raman spectra, as obtained for three different excitation lines at 633, 442 and 325 nm. All spectra present a number of bands, varying in intensity and sharpness. It is a remarkable feature of the Raman spectra in Figure 2.a that the obtained signatures depend significantly on the excitation wavelength. This change can be understood by three different (but linked) factors:

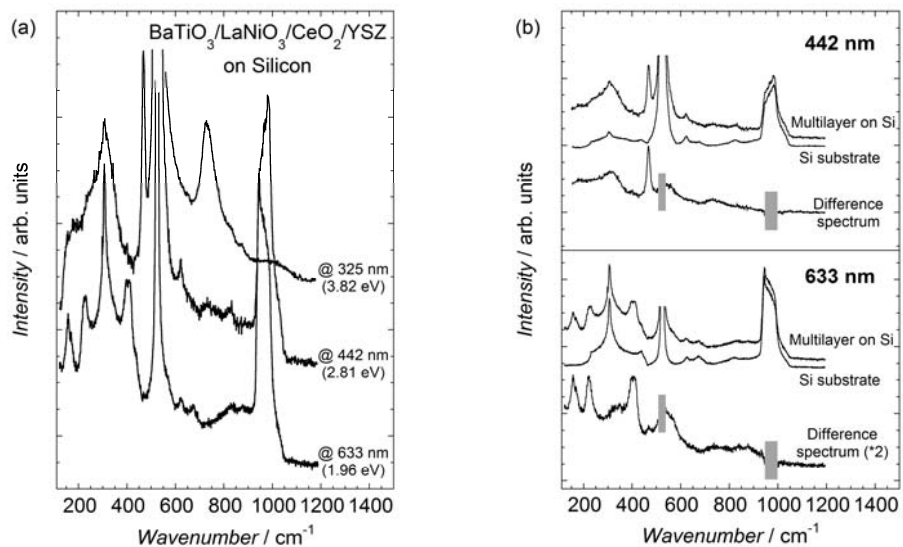

**Figure 2**

(a) Comparison of Raman spectra obtained at 325, 442 and 633 nm for the BaTiO$_3$/LaNiO$_3$/CeO$_2$/YSZ//Si(001) heterostructure. (b) Illustration of the normalisation and subtraction of the Si-substrate spectrum, described in the text. The gray boxes mask deformed spectral features from the subtraction of the strongest Si scattering.

(*i*) *Optical confocal depth*. Even though the exact experimental determination of the confocal depth is complex (as it depends both on the material and experimental set-up) simple optics implies that the optical depth for a given material decreases with decreasing wavelength.

(*ii*) *Absorption*. Generally speaking, a decrease in the laser wavelength can lead to photon energies above the band gap, which in turn increases the absorption and thus reduces the penetration depth. The best quantitatively studied illustration is silicon, where it has been shown that the penetration depth goes from 300 nm in the case of a 485 nm laser to only ~ 15 nm for a UV 364 nm laser.[32] This reasoning holds also for functional oxides: i.e. it has been shown that UV-Raman spectroscopy is a useful tool for probing nanoscale ultrathin films of ferroelectric oxides, such as $BaTiO_3$, on the top of a strongly scattering substrate.[14-15]

(*iii*) *Raman cross section*. It is well known that the Raman intensity of a given material can depend on the exciting wavelength. This situation is particularly pronounced when the incoming laser excitation is close to energies near the band gap. Depending on the specific energy, this can lead to pre-resonance effects (typical enhancement by a factor of 5 or 10) or even stronger enhancements up to $10^5$ in full resonance conditions. Raman resonant effects have been mainly studied on classic semiconductors like CdS (ref. [33]) or GaP (ref. [34]), while only little work has been reported on complex oxides, despite some promising reports.[35]

We will see later that all of the aforementioned phenomena have to be considered to understand the different Raman responses observed in these multilayers. In order to facilitate the discussion of the wavelength-dependent Raman signatures, it is useful to first separate the signal of the multilayer from the complex scattering signature of the silicon substrate. For this, we have collected Raman spectra of a bare Si substrate at the three different wavelengths, which we have then subtracted from the multilayer structure spectra, after a normalisation of the spectra to the Si-feature around 950 cm$^{-1}$. Figure 2.b illustrates this procedure for the 422 nm and 633 nm spectra together with the so-obtained Raman signature for the multilayer only. For the UV spectrum this subtraction is not necessary as the signal from the Si substrate is strongly reduced and does not interfere with the observed features.

In order to assign the different spectral features, Figure 3 compares the wavelength-dependent Raman spectra of the multilayer with Raman reference spectra of the pure individual constituents ($CeO_2$ and $BaTiO_3$ powders and a thick polycrystalline $LaNiO_3$ film deposited on bare Si), which we successively discuss going from the bottom to the top layer.

(i) *YSZ*: The signature of the bottom YSZ layer appears in none of the multilayer spectra. This is understood by considering literature reports of a very low Raman signal even for an uncovered thin film of cubic YSZ deposited on Si.[36]

(ii) *$CeO_2$*: In agreement with literature[37], the $CeO_2$ reference powder is characterized by a single $F_{2g}$ mode at around 465 cm$^{-1}$, while the second order features of $CeO_2$ are of much lower intensity. For the multilayer the $F_{2g}$ mode is observed as a sharp and strong mode in the 442 nm spectrum, while its intensity is strongly reduced in the 633 nm spectrum. Strong Raman spectra of $CeO_2$ under a 488 nm illumination have also been observed in the literature.[37] Our 442 nm is even closer to the reported[38] optical band gap suggesting that the strongly enhanced signature for the blue laser is related to pre-resonance effects, which are absent for the 633 nm.

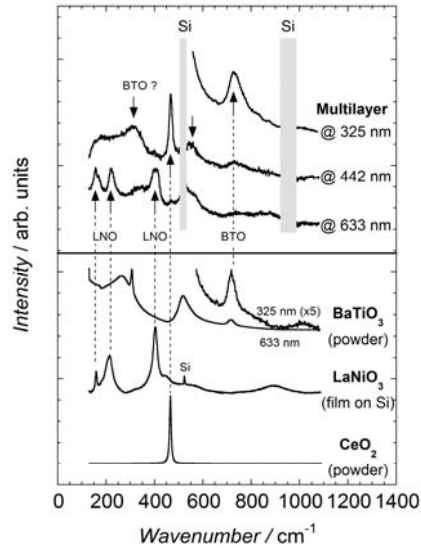

**Figure 3**

Comparison of Raman spectra obtained for the multilayer at 325, 442 and 633 nm (top panel, after subtraction of the Si-substrate signal) with reference spectra of the individual layers in the multilayer (bottom panel): BaTiO$_3$ (BTO), LaNiO$_3$ (LNO) and CeO$_2$ powders and a thick polycrystalline LaNiO$_3$ film deposited on bare Si). The vertical lines follow the band assignment discussed in the text. The gray boxes mask deformed spectral features from the subtraction of the strongest Si scattering.

(iii) *LaNiO$_3$*: LaNiO$_3$ crystallizes in a rhombohedrally distorted perovskite structure with space group $R\bar{3}c$, described by anti-phase tilts of the adjacent NiO$_6$ octahedra about the [111]$_p$ pseudo-cubic diagonal, described by the $a^-a^-a^-$ tilt system in Glazer's notation.[39] The reference spectrum of LaNiO$_3$ on a silicon substrate is characterized by three pronounced Raman bands around 155, 210 and 400 cm$^{-1}$ (ref. [40]). While such features are mostly invisible in the 442 nm spectrum of the multilayer, they are well distinguished in the 633 nm spectrum, in agreement with earlier work reporting an increased Raman response at this wavelength.[40]

(iv) *BaTiO$_3$*: The powder Raman spectrum of BaTiO$_3$ shows in agreement with literature[41-42] three dominating bands of $A_1$ symmetry at around 265, 515 und 715 cm$^{-1}$, labelled respectively TO2, TO3 and LO3. This characteristic signature can be only tentatively observed in the multilayer spectra obtained at 442 and 633 nm, but does not provide a convincing or analysable signature. In sharp contrast to this, the multilayer spectrum obtained with a UV 325 nm excitation line shows a well-defined signature of the high-wavenumber LO3 band of BaTiO$_3$. Such a well-defined signature is even the more striking since it is the weakest of the bulk characteristic BaTiO$_3$ signatures. This observation is understood by considering that the incoming UV phonon energy (3.92 eV) is above the band gap of BaTiO$_3$, leading to an increase in absorption and thus a reduced penetration depth, which in turn allows probing directly BaTiO$_3$, without being masked by the background scattering of the other layers. This observation confirms earlier reports stating that UV Raman spectroscopy is a well-adapted technique for observing signals from even ultrathin BaTiO$_3$ films.[14-15]

All in all, we observe that all constituents of the multilayer system, with the exception of YSZ, are observed in the Raman signature and, conversely, all features in the three multilayer spectra can be assigned to the individual layers or the silicon substrate. While all spectra contain the Si substrate signal, the 633 nm spectrum is dominated by LaNiO$_3$, the 442 nm spectrum by CeO$_2$ and slight traces only of LaNiO$_3$ and BaTiO$_3$ and, finally, the UV spectrum above 500 cm$^{-1}$ by a distinct signature of the BaTiO$_3$ layer.

## 5. Strain states in the multilayer

After the assignment of the different modes in the multilayer spectrum it is now interesting to inspect the individual modes in more detail to discuss potential strain-induced changes. Generally speaking, strain in thin films manifests in a Raman spectrum mainly via shifts in position of the Raman bands with respect to a strain-free reference material. Most commonly, compressive strain leads to an increase in wavenumber, while the opposite is observed for tensile strain. Based on this and the use of reference data, the strain state can be estimated, as will be discussed hereafter for every layer.

The BaTiO$_3$ and CeO$_2$ powders and the thick polycrystalline LaNiO$_3$ film on Si are all considered to be strain-free, thus serving as a reference for the discussion of strain effects.

Figure 4.a compares the $F_{2g}$ mode for the CeO$_2$ powder reference sample and the 442 nm spectrum of the multilayer. It can be seen that the CeO$_2$ mode of the multilayer is shifted by ~ 1.7 cm$^{-1}$ to higher wavenumber and while the peak width is increased. We believe that these spectral changes are not dominated by a nano-size effect of the 23 nm thick CeO$_2$ layer, as this would lead to a lowering of the wavenumber.[43] Rather, we attribute the positive wavenumber shift to a compressive strain state. Although the strain state of the thin film cannot be directly compared to hydrostatic pressure data, we note that high-pressure studies[43] indicate that the $F_{2g}$ mode shifts under a compressive deformation at a rate of + 3.3 cm$^{-1}$/GPa,

according to which the strain of $CeO_2$ in the multilayer structure can be estimated to ~ 0.5 GPa. The increased width indicates a lower coherence length in the multilayer, as often observed in thin films.

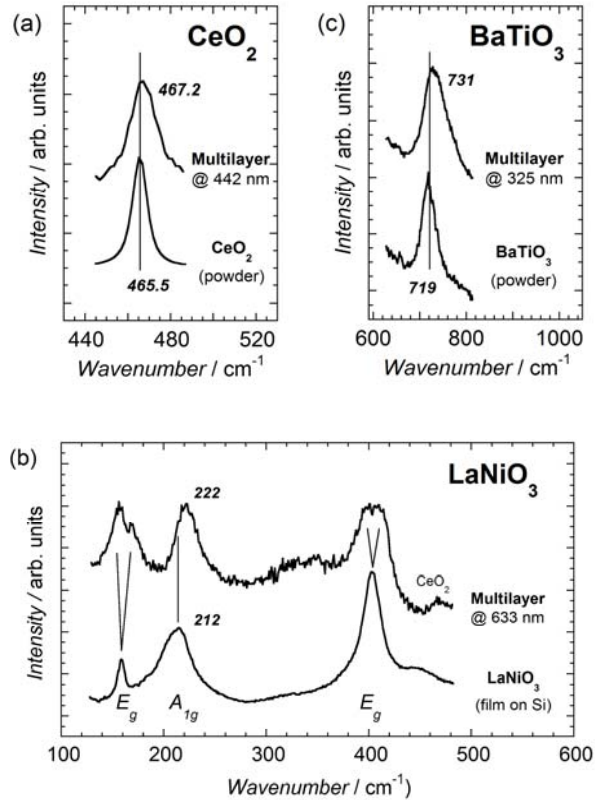

**Figure 4**

Comparison of spectral features in the multilayer with reference spectra (a) $F_{2g}$ mode of a $CeO_2$ powder and the 442 nm spectrum of the multilayer. (b) $LaNiO_3$ film on Si and the 633 nm spectrum of the multilayer. The vertical lines follow the band assignment discussed in the text. (c) $A_{1g}$ (LO3) mode of $BaTiO_3$ for a reference powder sample and the 325 nm spectrum of the multilayer.

Figure 4.b then compares the Raman spectrum of the relaxed $LaNiO_3$ reference film to the 633 nm spectrum of the multilayer. Two signatures can be distinguished for the multilayer: first, a strong +10 cm$^{-1}$ shift of the $A_{1g}$ mode and, second, a splitting of the two $E_g$ modes around 155 and 400 cm$^{-1}$. Both signatures provide direct evidence for significant strain-induced changes. The $A_{1g}$ mode of $LaNiO_3$ has been earlier assigned to a soft mode, the vibration pattern of which is described by rotations of the $NiO_6$ octahedra.[40, 44] As a consequence, this mode can be used in strained $LaNiO_3$ thin films for probing changes in the octahedra rotation angle, and thus the deviation from the ideal cubic phase.[40] Based on the experimental[40] and theoretical[44] observation that the $A_{1g}$ mode position scales by ~ 23 cm$^-$

$^1$/deg with the tilt angle, our observed shift suggests that the NiO$_6$ octahedra tilt angle of the embedded 30 nm thin LaNiO$_3$ layer is increased by 0.5° with respect to the bulk sample. In addition to this, the doubling of the two $E_g$ modes around 155 and 400 cm$^{-1}$ provides conclusive evidence that the thin film strain does not only act on the distortion, but induces a symmetry breaking towards a new crystal structure. Such a symmetry breaking is understood by recalling that the LaNiO$_3$ film is oriented along the pseudo-cubic [001]$_{pc}$ direction normal to the substrate, so that the bi-axial strain deformation acts in the basal (001)$_{pc}$ plane. For symmetry reasons, the octahedra tilt system $a^-a^-a^-$ of the rhombohedral $R\bar{3}c$ structure of LaNiO$_3$ cannot accommodate such a strain in the basal (001)$_{pc}$ plane but will adopt, for instance, a monoclinic structure. One possible candidate for the structure of the strained embedded layer is the monoclinic space group $C2/c$ described by the tilt system $a^-b^-b^-$, which allows maintaining the out-of plane $a^-$ tilt while the two $b^-$ tilts accommodate the strain. The very same space group has been recently proposed for strain accommodation in single layer LaNiO$_3$ thin films on (001) perovskite-type substrates.[45] We also note that the here observed strain-induced phase transition in LaNiO$_3$ is reminiscent of recent Raman observation in the multiferroic single layers of rhombohedral BiFeO$_3$ on a (001)LaAlO$_3$ substrate.[17]

Finally, Figure 4.c presents a comparison of the $A_{1g}$ (LO3) Raman mode of BaTiO$_3$ for a reference powder sample and the 325 nm spectrum of the multilayer. The LO3 mode is shifted by +12 cm$^{-1}$ in the multilayer, indicating a compressive strain state. Following high-pressure experiments in the literature, the typical compression deformation shift of the TiO$_6$ high-wavenumber in BaTiO$_3$ and other titanates is about 4 to 5 cm$^{-1}$/GPa (ref. [46-47]) so that the observed 12 cm$^{-1}$ shift suggests a significant compressive strain of the order of 2.5 to 3 GPa, in agreement with the relatively large bulk lattice mismatch between BaTiO$_3$ (a$_T$ = 3.994 Å) and the underlying LaNiO$_3$ (a$_{pc}$ =3.81 Å). Furthermore, it is known that the LO3 mode increases in wavenumber from 712 to 719 cm$^{-1}$ when going through the temperature-induced rhombohedral-orthorhombic-tetragonal phase sequence.[41] Based on this, we conclude that the BaTiO$_3$ layer at the top of the multilayer adopts a tetragonal structure, where the tetragonality is increased with respect to the bulk due to in-plane compressive strain.

## 5. Conclusion

In summary, we have presented a Raman scattering investigation of a complex BaTiO$_3$/LaNiO$_3$/CeO$_2$/YSZ//Si multilayer. It is shown that the Raman signature of the multilayer changes significantly for three different laser wavelength (633, 442 and 325 nm), which is explained by the wavelength-dependence of the optical depth, absorption and Raman cross section for the individual layers. While the use of a single wavelength does not allow a full investigation of the multilayer, our results demonstrates that Raman scattering at various wavelengths allows identification and analysis of the individual layers of functional oxide multilayers.

It is shown that all layers of the investigated multilayer are strained with respect to the bulk reference samples with a strain state that can be estimated via the Raman band positions. Interestingly, our results show that interfacial strain induces for the embedded $LaNiO_3$ even a new crystal structure away from its rhombohedral bulk structure. Such fine structural details of individual layers in a complex multilayer are difficult to obtain by other techniques, which further underlines the versatility of Raman spectroscopy for multifunctional oxides; be it in bulk, thin film or—as shown here—in complex heterostructures.

Generally speaking, we suggest that Raman scattering at various wavelengths offers a well-adapted non-destructive probe for the investigation of strain and structural changes in complex multilayer thin film structures and likely also in more complex 3D heterostructure or even nano-composites.

*Acknowledgements*

JK acknowledges financial support during his sabbatical stay at Warwick University from the Institute of Advanced Study (IAS) Warwick and from the Région Rhône-Alpes (CMIRA grant). Some of the equipment used at the University of Warwick was obtained through the Science City Advanced Materials project "Creating and characterizing Next Generation Advanced Materials". Financial support by the Spanish Government (Projects MAT2011-29269-C03 and CSD2007-00041) and by the Generalitat de Catalunya (Grant No. 2009-SGR-00376) is acknowledged.